\documentclass[12pt]{iopart}
\begin{document}
\def\beqn{\begin{eqnarray}}
\def\eeqn{\end{eqnarray}}

\title{Self-consistency in Theories with a Minimal Length}
 
\author{S.~Hossenfelder}

\address{Department of Physics, University of California,
Santa Barbara, CA 93106-9530, USA}
\ead{sabine@physics.ucsb.edu}

\begin{abstract}
 The aim of this paper is to clarify the relation between three 
different approaches of theories with a minimal 
length scale: A modification of the Lorentz-group in the 'Deformed Special Relativity', 
theories with a 'Generalized Uncertainty Principle' and those with 'Modified Dispersion Relations'.
It is shown that the first two are equivalent, how they can be translated into each other, 
and how the third can be obtained from them. An adequate theory with a minimal length scale requires
all three features to be present.

\end{abstract}


\section{The Role of the Planck Scale}	

Gravity itself is inconsistent with physics
at very short scales.
The introduction of gravity into quantum field theory appears to spoil their renormalizability and 
leads to incurable divergences. It has therefore been suggested that gravity
should lead to an effective cutoff in the ultraviolet, i.e. to a minimal observable length.
It is amazing enough that all attempts 
towards a fundamental theory imply the existence 
of such a minimal length scale. It is expected that the minimal length, $L_{\rm m}$ is close by, or
identical to the Planck length.

Motivations for the occurrence of a minimal length are manifold. 
A minimal length can be found in String Theory \cite{Gross:1987ar,Konishi:1989wk,Amati,Yoneya:1989ai}, 
Quantum Loop Gravity \cite{Rovelli:1997yv,Thiemann:2002nj,Perez:2003vx,Ash}, 
and Non-Commutative Geometries \cite{Douglas:2001ba,Girelli:2004md}.
It can be derived from various studies of thought-experiments \cite{Kuo:1993if,Daftardar:1986ci,quantum,Ng:1993jb}, 
phenomenological examinations of precision measurements \cite{Wigner,Salecker,Barrow,Sasakura:1999xp},
from black hole physics \cite{Padmanabhan:1986hd,Scardigli:1999jh}, the holographic
principle \cite{Ng:2004bt}, a T-duality of the path-integral \cite{Padmanabhan:1996ap,Smailagic:2003hm,Fontanini:2005ik} 
and probably further more.  For reviews, the interested reader is referred to 
\cite{Garay:1994en,Ng:2003jk,Hossenfelder:2004gj}.
The listed points are cross-related in many ways. The examination of these similarities
in the presently availably approaches is a promising way to increase our knowledge about the 
quantum nature of gravity.

Besides the various attempts to pin down the emergence of a finite resolution of spacetime,
the inclusion of the minimal length into the theoretical framework of the Standard Model (SM) has been 
examined from different sides: The Generalized Uncertainty Principle (GUP), 
Deformed Special Relativity (DSR) and Modified Dispersion Relations (MDR). These theories are an
effective description of the expected effects of quantum gravity and provide us with a useful
framework to describe the phenomenology of physics beyond the SM. 

The aim of this work is
to clarify the interrelationships of the different approaches to ensure the 
self-consistency of theories with a fundamental minimal length scale.  

This paper is organized as
follows. In the next section, we will briefly introduce the basic formalism of the three approaches. In 
section three we will examine the relations between them. We conclude in section \ref{concl}. Throughout
this paper we use the convention $\hbar=c=1$. 

\section{The Three Faces of the Minimal Length}

The phenomenology that arises from a finite resolution of space-time, 
and the mathematical structure associated with it, have been 
investigated closely \cite{Kempf:1994su,mlg,Maggiore:1993zu,Camacho,hydrogen,Brau,Akhoury:2003kc}. 
In the scenario without extra dimensions, the derived modifications are 
important mainly for structure formation
and the early universe \cite{earlyuniverse,Shankaranarayanan:2002ax,Mersini:2001su,Kempf:2000ac,Kempf:2001fa,Martin:2000xs,Easther:2001fz,Brandenberger:2000wr}. 
The importance to deal with the minimal length is sensibly enhanced if we consider a spacetime with 
large extra dimensions \cite{Hossenfelder:2003jz,Hossenfelder:2004up,Bhattacharyya:2004dy}.

\subsection{GUP} 

Test particles of a sufficiently high energy to resolve a distance as small as the Planck length
are predicted to gravitationally curve and thereby to significantly disturb the structure of the
spacetime which they are meant to probe. Thus, in addition to the expected quantum uncertainty, 
there is another uncertainty caused which arises from spacetime fluctuations at the Planck scale.

This behavior can be quantified by allowing the properties of the wave-vector 
${\bf k} = (\omega, \vec{k})$ to be modified at highest energies, such that ${\bf k}$ is
no longer linear to the momentum ${\bf p}$. In particular, we will 
want the wave-length and thereby the resolution of spacetime to have the lower bound $L_{\rm m}$, no
matter how much we increase ${\bf p}$. 
 
In the following, we will restrict ourselves to the isotropic case in which it will be 
sufficient to work with one space-like dimension.
We denote the simplified wave-vector as ${\bf k} = (\omega,k)$.
Denoting the momentum  with
${\bf p} = (E,p)$, we can quantify the modifications by $f = (f_0,f_1)$, where
\beqn
f_0 ({\bf p}) = \omega  \quad,\quad  f_1 ({\bf p}) = k \quad. \label{f}
\eeqn
We will assume that the function is well-defined 
in a suitable manner, smooth and differentiable and that it is a one-to-one map which can be inverted
\beqn
(f^{-1})_0 ({\bf k}) = E  \quad,\quad  (f^{-1})_1 ({\bf k}) = p \quad. \label{f-1}
\eeqn
It has to fulfill the low energy limit
\beqn
f = f^{-1} = {\rm Id} \quad {\mbox{for}} \quad \omega,|k| \ll 1/L_{\rm m} \quad, \label{cons1}
\eeqn
and it should be bounded by the minimal length:
\beqn
|f_0|, |f_1| \leq 1/L_{\rm m} \quad {\mbox{for all}} \quad {\bf p} \quad. \label{cons2}
\eeqn

The quantization of this {\sl ansatz} is straightforward and follows the usual procedure. 
The commutators between the corresponding operators $\hat{k}$ and $\hat{x}$ 
remain in the standard form. 
Using the well known commutation relations and inserting the functional relation between the
wave vector and the momentum then yields the modified commutator for the momentum 
\begin{eqnarray}  
[\hat x,\hat k]={\mathrm i } \quad\Rightarrow\quad 
[\hat{x},\hat{p}]&=& {\rm i} \partial_k (f^{-1})_1({\bf k}) \quad.
\end{eqnarray}
This results in the Generalized Uncertainty Principle ({\sc GUP)}
\begin{eqnarray} \label{gu}
\Delta p \Delta x \geq \frac{1}{2}  \big| \left\langle \partial_k (f^{-1})_1({\bf k}) 
\right\rangle \big| \quad, 
\end{eqnarray}
which reflects the fact that by construction it is not possible any more to resolve space-time distances
arbitrarily well. Since $f_1({\bf p})$ gets asymptotically constant, the derivative $\partial_p f_1$
drops to zero and the uncertainty in Eq. (\ref{gu}) increases for large momenta. 
We will refer to this theory as a GUP, if $f \neq$ Id.

Various examples for the function $f$ can be found in \cite{Unruh:1994je,Magueijo:2002am,Cortes:2004qn}. 
A very common choice is
$f_0({\bf p}) = g(E)$ ,  $f_1({\bf p}) = g(p)$ 
with
\beqn
g(x) = \tanh(L_{\rm m} x) \quad. \label{ex1}
\eeqn
For many applications, only a first order expansion of $f$ is examined, in which
case the constraint Eq. (\ref{cons2}) might not be fulfilled for all ${\bf p}$. For example,
 consider the case  \cite{Amelino-Camelia:2000mn}
\beqn
f_0({\bf p}) = E \quad, \quad f_1 ({\bf p}) = p (1  + \alpha L_{\rm m}	~  E) \quad, \label{ex2}
\eeqn
where $\alpha$ is some constant parameter of order one.

\subsection{DSR}

By definition, a minimal length should not undergo a Lorentz-contraction when it is
boosted. That means, a modification of the Lorentz-transformations becomes necessary.
The new transformations should not only leave the speed of light invariant, but have the minimal
length as a second invariant. Such a modification can be achieved without introducing 
exceptional reference frames by a deformation of the usual 
Lorentz-tranformation \cite{Magueijo:2002am,Cortes:2004qn,Amelino-Camelia:2000mn,Amelino-Camelia:2002wr,Sasakura:2000vc,Magueijo:2001cr,Toller:2003tz,Rovelli:2002vp,Deriglazov:2004yr,Deriglazov:2005vd}. 
These transformations can be described through the generators of the 
$\kappa$-Poincar\'e Hopf algebra \cite{Lukierski:1991pn,Majid:1994cy,Lukierski:1993wx,Kowalski-Glikman:2001gp,Bruno:2001mw}, 
the exact relation of which to possible {\sc DSR}-theories has been 
investigated in \cite{Kowalski-Glikman:2002we}.

The momentum is a Lorentz-vector in the standard way, and it transforms  according to the usual 
Lorentz-transformation, $\Lambda$. The matrix $\Lambda$ is an element of the Lorentz-group $SO(3,1)$, 
and can be parameterized by the six parameters of the group. These 
parameters will encode the nature
of the transformation, i.e. the type of rotation and the boost. 
Under a change of inertial systems
the transformation is then 
${\bf p}' = \Lambda {\bf p}$.

In order to enable the invariance of the minimal length, it is now assumed 
that the wave vector behaves according to an unknown new transformation $\widetilde{\Lambda}$.
The modified transformation property of the wave-vectors are thereby achieved by allowing the generators
of the Lorentz-group to act non-linearly on the space of wave-vectors. Exponentiating the 
infinitesimal transformations then results in an explicit dependence of the modified Lorentz-transformation on the
wave-vector. We will denote this as 
${\bf k}' = \widetilde\Lambda({\bf k})$,
where the transformations $\widetilde\Lambda$ fulfill the requirements of forming a group.

The possible modification of the Lorentz-transformation at highest energies has recently received
large interest as a candidate to explain the observations of ultra high energetic cosmic rays 
\cite{UHECRs,Amelino-Camelia:2002gv,Sarkar:2002mg,Chisholm:2003bu,Moffat:2002nu,Gonzalez-Mestres:2000eg,Bertolami:1999da,Amelino-Camelia:2000zs}.

\subsection{MDR}

The ordinary relativistic dispersion relation for a particle of mass $m$ has the form
\beqn
E^2 - p^2 - m^2 = 0\quad. \label{umdr} 
\eeqn
We will call the dispersion relation for ${\bf k}$ modified, when it takes the form
\begin{eqnarray}
\omega^2 - k^2 - m^2 = \Pi({\bf k}) \quad \mbox{with} \quad \Pi({\bf k}) \neq 0 \quad. \label{mdr} 
\end{eqnarray}
Confusingly, the term 'dispersion relation' is also widely used for the derivative 
$d k/d \omega$. 
In the
following, we will refer to the dispersion relation in the form of Eq. (\ref{umdr}). In the case in
which 
\beqn
\frac{d \omega}{d k} \neq \frac{d E}{d p} \quad, 
\eeqn
we will have a theory with a variable speed of light (VSL). As will be discussed below, a MDR needs not
necessarily imply a VSL.

\section{Relations Between Different Approaches}

It is apparent that the three different approaches are related to each
other and that they must be treated as a threesome for a self-consistent framework. 
Even though the presence of relationships in special cases has been examined 
previously \cite{Cortes:2004qn,Calisto:2005mz}, a clarification of the 
precise form remains to be given. 
It is of particular interest, whether each {\sl ansatz} is equivalent to the others,
and if so, how the one can be obtained from the other in practice.

\subsection{GUP $\Rightarrow$ DSR}

This GUP is connected with the DSR in a very general way by observing that once the functional relation $f$
between the quantities ${\bf k}$ and ${\bf p}$ is known, the transformation of ${\bf k}$ can be obtained
from that of ${\bf p}$. Even though the wave-vector transforms in an unknown way,
we can  find it by using the related momentum with help of $f$, then 
applying the standard Lorentz transformation to the momentum, and finally using the inverse of $f$:
\beqn
\widetilde{\Lambda}({\bf k}) := f( \Lambda f^{-1}({\bf k} )) \quad. \label{gupdsr}
\eeqn

\subsection{DSR $\Rightarrow$ GUP}

Now let us assume that we know the new transformation $\widetilde{\Lambda}$ for the
wave-vector ${\bf k}$ in addition to the standard transformation $\Lambda {\bf p}$, 
and we aim to know the pair $(f({\bf p}),{\bf p})$ for all 
values of ${\bf p}$. Therefore, let us first remember that we know some special pairs 
$(f({\bf p}'),{\bf p'})$ 
in the low energy regime (rest frame or very red shifted), 
where $|E'|, |p'| \ll m_p$. In this limit, we will have no modifications 
\beqn
{\bf k'} = {\bf p'} \quad{\mbox{and so}} \quad f = {\rm Id} \quad.
\eeqn
We then obtain the function $f$ at all energies by boosting it into a new inertial system in which ${\bf p}$ 
takes an arbitrary value. Since this boost is known, we can define the relation $f$ for all ${\bf p}$ via
\beqn
f({\bf p}) := \widetilde{\Lambda}({\rm Id}(\Lambda^{-1} {\bf p} )) \quad.
\eeqn

\subsection{DSR,GUP $\leftrightarrow$ MDR}

Now that we have seen how GUP is equivalent to DSR, let us examine their relation to the MDR. From the
previous arguments, it will be sufficient to examine the way the GUP and MDR do affect each other. 
From a simple counting of equations one can already see that both in general will not be equivalent. 
The MDR in Eq. (\ref{mdr}) is
one equation to relate the pairs of $(\omega,k)$, but we will not know
how these are related
to the pairs of $(E,p)$. Even with the assumption of isotropy, recovering 
GUP requires the knowledge of two unknown
functions $f_0$ and $f_1$. 

Using the relation between ${\bf p}$ to ${\bf k}$ one can, however, immediately write down the 
form of the dispersion relation with a GUP 
\beqn
((f^{-1})_0({\bf k}))^2 - ((f^{-1})_1({\bf k}) )^2 -m^2 \label{gupmdr}
&=& 0 \quad.
\eeqn
Comparing with Eq. (\ref{mdr}), we find the translation of GUP into MDR
\beqn
\Pi({\bf k}) &=& \omega^2  - ((f^{-1})_0({\bf k}) )^2 - k^2 + ((f^{-1})_0({\bf k}))^2 
\quad. \label{pi}
\eeqn

Even if it will, in general, not be possible to obtain the GUP-functions out of the MDR, there is an 
important and frequently used case in which there exists a useful relation:
When both components of the vector do not mix, that is they take the form
$f_0({\bf p}) = f_0(E)$ and $f_1({\bf p}) = f_1(p)$, or 
\beqn
(f^{-1})_0({\bf k}) = (f^{-1})_0(\omega) \quad,\quad (f^{-1})_1({\bf k}) = (f^{-1})_1(k)\quad, \label{case1}
\eeqn 
respectively (the example in Eq. (\ref{ex1}) is of this case) then 
$\Pi({\bf k})$ takes the special form 
\beqn
\Pi({\bf k}) = \widetilde\Pi_0({\omega}) - \widetilde\Pi_1({k}) \quad. \label{sep}
\eeqn
Since $\Pi(0)=0$, Eq. (\ref{pi}) can then be cleanly separated and yields the invertible relations
\beqn
(f^{-1})_{0/1}(\omega)    &=& \left( \widetilde\Pi_{0/1}({\omega}) - 1 \right)^{1/2}
 \quad.
\eeqn

\subsection{MDR $\leftrightarrow$ VSL}

As mentioned earlier, a MDR needs not necessarily imply a varying speed of light. Using Eq.(\ref{gupmdr}) with
zero rest mass and taking the square root results in
\beqn
(f^{-1})_0({\bf k}) = (f^{-1})_0({\bf k}) \quad, \label{bmdr}
\eeqn
which implicitly defines the mass-shell condition as $\omega(k)$. The total 
derivative with respect to $k$ yields 
\beqn
\partial_\omega \left( (f^{-1})_0 - (f^{-1})_1 \right) = 
\partial_k \left( (f^{-1})_0 - (f^{-1})_1 \right) \quad, \label{VSL}
\eeqn
taken at the position ${\bf k} = (\omega(k), k)$. Therefore, the theory will leave the
speed if light unmodified if $f$ fulfills the constraint
\beqn
(f^{-1})_0(x,x) - (f^{-1})_1(x,x) =0 \quad,
\eeqn
which expresses the invariance of the light-cone. Two useful classes of function that
respect the constancy of the speed of light, and which are appealing because of their 
symmetry are
\beqn
f^{-1}({\bf k}) &=& {\bf k}~ g({\bf k}) \quad \quad \quad \quad \quad~ {\mbox{with}} \quad g(0) \to 1 \label{vsl1} \\
f^{-1}({\bf k}) &=& (h(\omega,k), h(k,\omega)) \quad {\mbox{with}} \quad h(0) \to x \quad, \label{vsl2}
\eeqn
and with otherwise arbitrary $g,h$. 
For these functions, the
dispersion relation for vanishing rest mass is not modified as one sees directly by inserting them in Eq.(\ref{bmdr}) 
from which follows that $\Pi({\bf k}) =0$. The dispersion relation in the form Eq.(\ref{mdr}) 
will 
be modified for massive particles, even though the speed of light is still $1$. 
Functions of the type $f^{-1}({\bf k}) = (g(\omega), g(k))$, which were 
discussed in \cite{Hossenfelder:2003jz} are a special case of (\ref{vsl2}). 
Functions of the type (\ref{vsl1}) were discussed in 
\cite{Magueijo:2002am}.

\begin{figure}[h]
\begin{center}
\unitlength 1cm
\begin{picture}(7,6)(0,1) 
\linethickness{.3mm}
\put (0,5){\framebox(2,1){\shortstack{GUP \\  ${\bf k} = f ({\bf p})$}}}
\put (5,5){\framebox(2,1){\shortstack{DSR \\  ${\bf k}' = \widetilde{\Lambda}({\bf k})$}}}
\put (1.4,2){\framebox(4.2,1){\shortstack{MDR \\  ~$\omega^2 - k^2 - m^2 = \Pi({\bf k})$}  }}
\put (1.4,.97){\framebox(2.07,1){\shortstack{w/o VSL \\  ~$d \omega /d k = 1$} }}
\put (3.5,.97){\framebox(2.1,1){\shortstack{with VSL \\  ~$d \omega /d k \neq 1$} }}

\put (2.5,5.4){\line(1,0){2}}
\put (4.5,5.6){\line(-1,0){2}}
\put (2.3,5.5){\line(1,1){0.4}}
\put (2.3,5.5){\line(1,-1){0.4}}
\put (4.7,5.5){\line(-1,1){0.4}}
\put (4.7,5.5){\line(-1,-1){0.4}}

\put (1,4.8){\line(1,-1){1.41}}
\put (2.5,3.3){\line(-1,0){0.41}}
\put (2.5,3.3){\line(0,1){0.41}}

\put (6,4.8){\line(-1,-1){1.41}}
\put (4.5,3.3){\line(1,0){0.41}}
\put (4.5,3.3){\line(0,1){0.41}}

\end{picture} 
\caption{Summary of the derived relations between the examined theories: Generalized Uncertainty Principle (GUP),
Deformed Special Relativity (DSR), Modified Dispersion Relation (MDR), Varying Speed of Light (VSL). \label{fig1}}
\end{center}

\vspace*{-.2cm}
\end{figure}
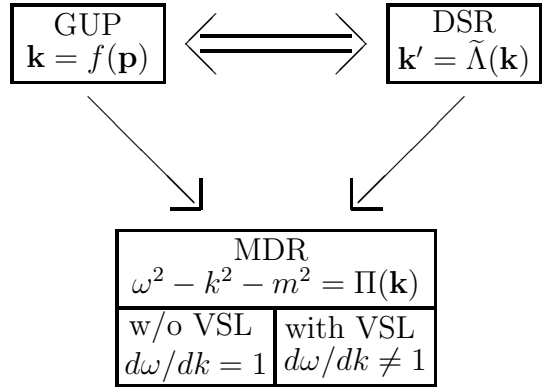

\section{Conclusions}
\label{concl}
We have shown that theories with a Generalized Uncertainty Principle are equivalent to
these with a Deformed Special Relativity and that they can be obtained from each other in
a straightforward way. We have derived how both result in a modified version of the 
dispersion relation, which needs not necessarily imply a varying speed of light. The explicit
translations between the existing approaches have been given. Provided that all three modifications
are made together, the framework is self-consistent and can be used to extend the Standard Model.

The found relations between the different approaches towards a theory with a minimal length scale are summarized
in Fig. \ref{fig1}. 

\section*{Acknowledgments}
 
This work was supported by  the {\sc DFG} and by the Department of Energy under 
Contract DE-FG02-91ER40618. I thank Steve Giddings for helpful discussions.

\section*{References}

{
 
}
\end{document}